\documentclass[namedreferences]{SolarPhysics}
\usepackage{epsfig}

\usepackage{graphicx}

\newcommand{\etal}{{\it et~al.}}
\newcommand{\be}{\begin{equation}}
\newcommand{\ee}{\end{equation}}
\newcommand{\beq}{\begin{eqnarray}}
\newcommand{\eeq}{\end{eqnarray}}

\newcommand{\aeta}[3]{  #1, {\it Astron. Astrophys.} {\bf  #2}, #3.}

\newcommand{\aspj}[3]{  #1, {\it Astrophys. J.} {\bf  #2}, #3.}

\newcommand{\sph}[3]{   #1, {\it Solar Phys.} {\bf  #2}, #3.}

\newcommand{\arcsec}{^{\prime \prime}}

\begin{document}
\begin{article}
\begin{opening}

   \title{Widespread Occurrence of Trenching Patterns
   in the Granulation Field: Evidence for Roll Convection?}

   \author{A.V. \surname{Getling}$^{1}$ and
           A.A. \surname{Buchnev}$^{2}$}

   \runningauthor{Getling and Buchnev}

   \runningtitle{Trenched Granulation Patterns and Roll Convection}

   \institute{$^{1}$Institute of Nuclear Physics, Lomonosov
              Moscow State University, 119992 Moscow, Russia
              \\ \email{A.Getling@mail.ru}
              $^2$Institute of Computational Mathematics
              and Mathematical Geophysics, 630090 Novosibirsk,
              Russia\\ \email{baa@ooi.sscc.ru}}

   \date{Received April 29, 2007; accepted September 14, 2007}

\begin{abstract}
Time-averaged series of granulation images are analysed using
COLIBRI, a purpose-adapted version of a code originally
developed to detect straight or curvilinear features in aerospace
images. The algorithm of image processing utilises a nonparametric
statistical criterion that identifies a straight-line segment as a
linear feature (lineament) if the photospheric brightness at a
certain distance from this line is on both sides stochastically
lower or higher than at the line itself. Curvilinear features can
be detected as chains of lineaments, using a criterion modified in
some way. Once the input parameters used by the algorithm are
properly adjusted, the algorithm highlights ``ridges'' and
``trenches'' in the relief of the brightness field, drawing white
and dark lanes. The most remarkable property of the trenching
patterns is a nearly-universally-present parallelism of ridges and
trenches. Since the material upflows are brighter than the
downflows, the alternating parallel light and dark lanes should
reflect the presence of roll convection in the subphotospheric
layers. If the numerous images processed by us are representative,
the patterns revealed suggest a widespread occurrence of roll
convection in the outer solar convection zone. In particular, the
roll systems could form the fine structure of larger-scale,
supergranular and/or mesogranular convection flows. Granules
appear to be overheated blobs of material that could develop in
convection rolls due to some instabilities of roll motion.
\end{abstract}

\keywords{Sun: photosphere, Sun: granulation}

\end{opening}

\section{Introduction}

Getling and Brandt \cite{gb1} reported that images of solar
granulation averaged over time intervals as long as, \textit{e.g.,} two hours,
are far from completely smeared but contain bright, granular-sized
blotches, which may form quasi-regular systems of concentric rings
or parallel strips --- ``ridges'' and ``trenches'' in the
brightness field --- on a meso- or supergranular scale. Getling
\cite{g06} implemented a detailed investigation of such long-lived patterns
and found that they do not appear unusual in images averaged over
one to two hour time intervals.

These systems resemble some specific types of the roll patterns
known from laboratory experiments on Rayleigh--B\'enard convection
and may reflect the fine structure of subphotospheric convection
cells. If the time variations of intensity are traced at the point
corresponding to a local intensity maximum in the averaged image
(near the centre of a light blotch) and at a nearby local-minimum
point, a remarkable pattern of correlations between these
variations can be noted. In some cases, the correlations are
periodic functions of the time lag or a tendency to
anticorrelation is observed. This fact supports our suggestion
(Getling, 2000) that granules are hot blobs of the solar plasma
carried by the convective circulation and that they can even
re-emerge on the photospheric surface.

Since the quasi-regular structures manifest themselves in
time-averaged images, they should be associated with a long-lived
component of the granulation field. Getling and Brandt \cite{gb1}
and Brandt and Getling \cite{bg} noted that the decrease in the
rms contrast of the averaged images with the averaging time [$t$] is
considerably slower compared to the statistical $t^{-1/2}$ law,
and this fact was regarded as a particular argument for the
presence of long-lived features.

In his comment on the original paper by Getling and Brandt
\cite{gb1}, Rast \cite{rast} suggested that the structures in the
granulation field are merely of a statistical nature and do not
reflect the structure of real flows. He constructed a
series of artificial random fields with some characteristic parameters
typical of solar granulation and found some similarities between
the properties of these fields and of the real granulation
patterns. On this basis, Rast raised doubts on the existence of
the long-lived component in the granulation dynamics. A detailed
discussion of Rast's criticism is given by Getling \cite{g06}
invoking the contrast-variation analysis (Brandt and Getling
2004). A number of counterarguments are presented and it is shown
that Rast's considerations cannot be applied to the real
granulation patterns.

As noted by Getling and Brandt \cite{gb1}, signatures of the prolonged persistence of some features in the granulation patterns have already been observed previously. Roudier \etal\ \cite{roud7} detected long-lived singularities~--- ``intergranular holes'' (dark features)~--- in the network of supergranular lanes. Such holes
were continuously observed for more than 45~minutes, and their diameters varied
from 0.24$\arcsec$ (180 km) to 0.45$\arcsec$ (330 km). Later, Hoek\-ze\-ma, Brandt, and Rutten \cite{hoek3} and Hoekzema and Brandt \cite{hoekbr} also studied similar features, which could be
observed for 2.5~hours in some cases.  Baudin, Molowny-Horas, and Koutchmy \cite{baud} attributed the blotchy appearance of a 109-minute-averaged granulation image to a sort of persistence of the granulation pattern; alternatively, this effect can be interpreted in terms of the recurrent emergence of granules at the same sites (Getling, 2006).

Some indications of a long-term spatial organisation in the granulation field have also been revealed in observations. Dialetis \etal\ \cite{dial} found that granules with longer lifetimes exhibit a tendency to form mesogranular-scaled clusters. Muller, Roudier, and Vigneau \cite{mul} also detected such clustering in the spatial arrangement of large granules; they emphasised a plausible relationship between the clusters and mesogranules. Roudier \etal\ \cite{roud5} reported their observations of a specific collective behaviour of families (``trees'') of fragmenting granules. Such families can persist for up to eight hours and appear to be related to mesogranular flows. An imprint of the supergranulation structure can also be traced in the granulation pattern (Baudin, Molowny-Horas, and Koutchmy, 1997).

Based on their analysis of pair correlations in the supergranular
and granular fields, Berrilli \etal\ \cite{bdm-str} reported
a finding that bears some similarity with ours. Specifically, the
probability of finding a target supergranule or granule
(identified by its barycentre) at a given distance from the
barycentre of a chosen reference supergranule or granule is an
oscillating function of this distance with a local amplitude
decreasing from some maximum (reached at a small distance). This
reflects a specific kind of order in the supergranulation and
granulation patterns, which is consistent with the presence of
concentric-ring structures.

It is remarkable that an effect of spatial ordering has also been detected on a supergranular scale. Lisle, Rast, and Toomre \cite{lrt} revealed a persistent alignment of supergranules in a
meridional direction reminiscent of the alignment of light blotches described by us.
There is also a parallelism in the
interpretation of the quasi-regularity in the granulation and
supergranulation fields: while we are inclined to interpret the
trenching patterns as the manifestation of a fine structure of the
supergranular flows, Lisle, Rast, and Toomre \cite{lrt} attribute the
alignment of supergranules to an ordering influence of giant
convection cells.

Here, we employ a specific
image-processing algorithm to analyse granulation patterns and
investigate their spatial organisation. In particular, the results of
our study show that the arrangement of granules may assume various forms, and the patterns of light blotches detectable in the averaged images have a common feature~--- a nearly universally present parallelism of alternating light and dark lanes. If the images analysed by us are representative, the trenching patterns could naturally be interpreted as manifestations of roll
convection entraining granules; this form of motion proves to be widespread in the upper subphotospheric layers.

\section{The Data}

We mainly deal with excellent images of the photosphere from the
well-known La~Palma series obtained by Brandt, Scharmer, and Simon
(see Simon {\it et~al.}, 1994) on 5 June 1993 using the Swedish
Vacuum Solar Telescope (La Palma, Canary Islands). Specifically,
we use a seven-hour subset of this series and a $43.5\times 43.5$~Mm$^2$
subfield ($480\times 480$ pixels of size 90.6~km ) of the original
images. The observed area was located not far from the disk
centre, and the images were produced by the telescope in the
10-nm-wide spectral band centred at a wavelength of 468~nm. The
resolution was typically no worse than about 0.5$\arcsec$, and the
frame cadence was 21.03 seconds.

Previously (Getling and Brandt, 2002;
Getling, 2006), we already described the principal elements of the
data acquisition and pre-processing techniques employed by Simon
{\it et~al.} \cite{simon}. Here, we only briefly recall that the
images were aligned, destretched, and cleaned from rapid intensity
variations by means of subsonic filtering. Furthermore, all of
them were normalised to the same value of the rms contrast, and
the residual large-scale intensity gradients were removed from
them.

We give here our primary attention to granulation fields averaged
over one to two hour intervals.

In addition to the data of the La Palma series, we use here an
averaged image from the subsonically-filtered version of the
45.5-hour series obtained using the SOHO MDI instrument in 1997, from
17 January 00:01 UT to 18 January 21:30 UT (see Shine, Simon, and Hurlburt, 2000). This series contains white-light images with a
resolution of about 1.2$\arcsec$\ taken at a one-minute interval. We
present here an enlarged $87 \times 70$~Mm$^2$ cutout ($200 \times
160$ pixels) of the selected image.

\section{Image Processing}

In our analysis, we use the
\textit{CO\/}ntours and \textit{LI\/}neaments in the
\textit{BRI\/}ghtness field (COLIBRI) code, a purpose-adapted version of
a code constructed previously at the Institute of Computational
Mathematics and Mathematical Geophysics (Novosibirsk, Russia) and
intended for the detection of linear and circular structures in
aerospace images (with this design, the code has been successfully
employed for several years). The code is an implementation of
certain statistical criteria for the detection of stretched
features (in our case, linear structures and contour elements)
against a random background. The underlying algorithm was
developed by Salov \cite{salov}.

The principal reason for using such techniques is as follows.
The random component of the processes under study clutters the
observational data (\textit{i.e.,} images) with random
variations, which mask the brightness differences between the
points of the object sought and of the background areas. Accordingly, reliable
criteria of detection of objects should be based on a probabilistic
(statistical) approach. The presence
of an object in the image manifests itself in the fact
that the random observable quantities in the object are \textit{stochastically}~--- to say more simply, \textit{systematically}~---
larger (or smaller) than in the background areas. The so-called nonparametric criteria, based on stochastic
comparisons of the observables inside and outside the hypothetical object, do not depend on the (unknown) distributions of the
observables and prove to be efficient for our purposes.

The COLIBRI code has two operation modes that enable the
detection of either lineaments or contours in photospheric images.

A lineament (linear element) is a stretched, nearly linear object,
blurred and nonuniform in brightness along its extent. To detect a
lineament, the algorithm analyses (almost) all its possible positions. For any trial position, to decide whether or not a lineament
is present,
the set of brightness values
taken at some points along the hypothetical lineament of length $l$
is compared with the two sets taken at the points located at a certain distance $\Delta$ on both sides of the lineament, at the normals
intersecting it at the originally chosen points. Specifically, the algorithm
checks the
hypotheses that either all the sets of measured quantities are
stochastically equal (no object is present) or, alternatively, the set of the
central brightness values is stochastically greater (smaller)
than the sets taken on both sides (an object is
present). The nonparametric criterion used to test these hypotheses (Salov,
1997) is based on computing the values of the Mann--Whitney statistics (Mann and Whitney, 1947; Whitney, 1951).

A somewhat different approach is used in the regime of detection
of contours, by which curvilinear features with \textit{a priori}
unknown shapes are meant. In this case, the algorithm constructs
the contours as chains of lineaments, but the criterion for their
detection has a different form. Again, sets of brightness values are
taken at normals to the trial lineament at a certain
distance $\Delta$ on both sides of it. To decide whether or not an object
is present, the algorithm checks the hypotheses that either the
sets of brightness values taken on the two sides are stochastically
equal (no object is present) or, alternatively, the values on one
side are stochastically larger then the values on the other side
(an object is present). In this case, the nonparametric criterion
used to test these hypotheses is also based on the computed values of the
Mann--Whitney statistics (Salov, 1997). In essence, a trial
lineament is assumed to be a really-present object provided the
brightness gradient across it has some stochastically-predominant
sign. In this respect, the detected contours very crudely correspond to
isohypses (horizontals) in a topographic contour map; however, they do not
depend on the fine topographic details of the brightness field
because the intensity values at a contour are not taken into
account in constructing this contour.

The distance $\Delta$, which is everywhere measured in
pixels, is an input parameter of the algorithm. Other
important parameters are the admissible probability of spurious
detection of objects [$p$] and the range of admissible lengths of
lineaments [$l$] (we also express these lengths in pixels). The determination of the actual probability of spurious detection is fairly sophisticated (Salov, 1997), and describing it is beyond the scope of this paper.

Because of the noisy appearance of the analysed images, the
detection procedure straightforwardly applied to an image may reveal
a large number of spurious details (by noise, we mean the
presence of a multitude of light blotches interspaced by darker
features, any of which, taken alone, is not representative from
the standpoint of seeking quasi-regular structures). To reduce the
noise level, a special filtering procedure was developed.
It consists of two steps. First, any lineament is removed if it is
completely covered by another one; at the same time, lineaments
that partially overlap one another are combined into one lineament.
Second, all lineaments that have no parallel neighbours (\textit{i.e.,} do not
belong to any bunch of lineaments) are also removed. The effect of filtering will be illustrated below (in Figure~\ref{fr65unfilt}).

\section{Results}

Our primary goal was to analyse structures in time-averaged series
of granulation images. In addition, we attempted to process
individual images, or snapshots. Signs of regularity can be
visually noted in both cases; however, the snapshots contain many
more fine details than do the averages, so that much more spurious
features emerge when snapshots are processed. Here, we restrict
ourselves to considering only the structures detected in averaged
images.

The number of lineaments and contours detected in granulation
images strongly depends on the chosen parameters of the
image-processing procedure. It should be kept in mind that,
generally, there is no combination of parameters that could be
considered a universal optimum, applicable in any case. The optimum
really depends on both the character of the analysed pattern and
on the objective of analysis in any particular
situation.

\begin{figure} 
\centering
 {\scriptsize \hspace{0.6cm} (a) \hspace{5.6cm} (b)}
 \includegraphics[width=0.48\textwidth]
 {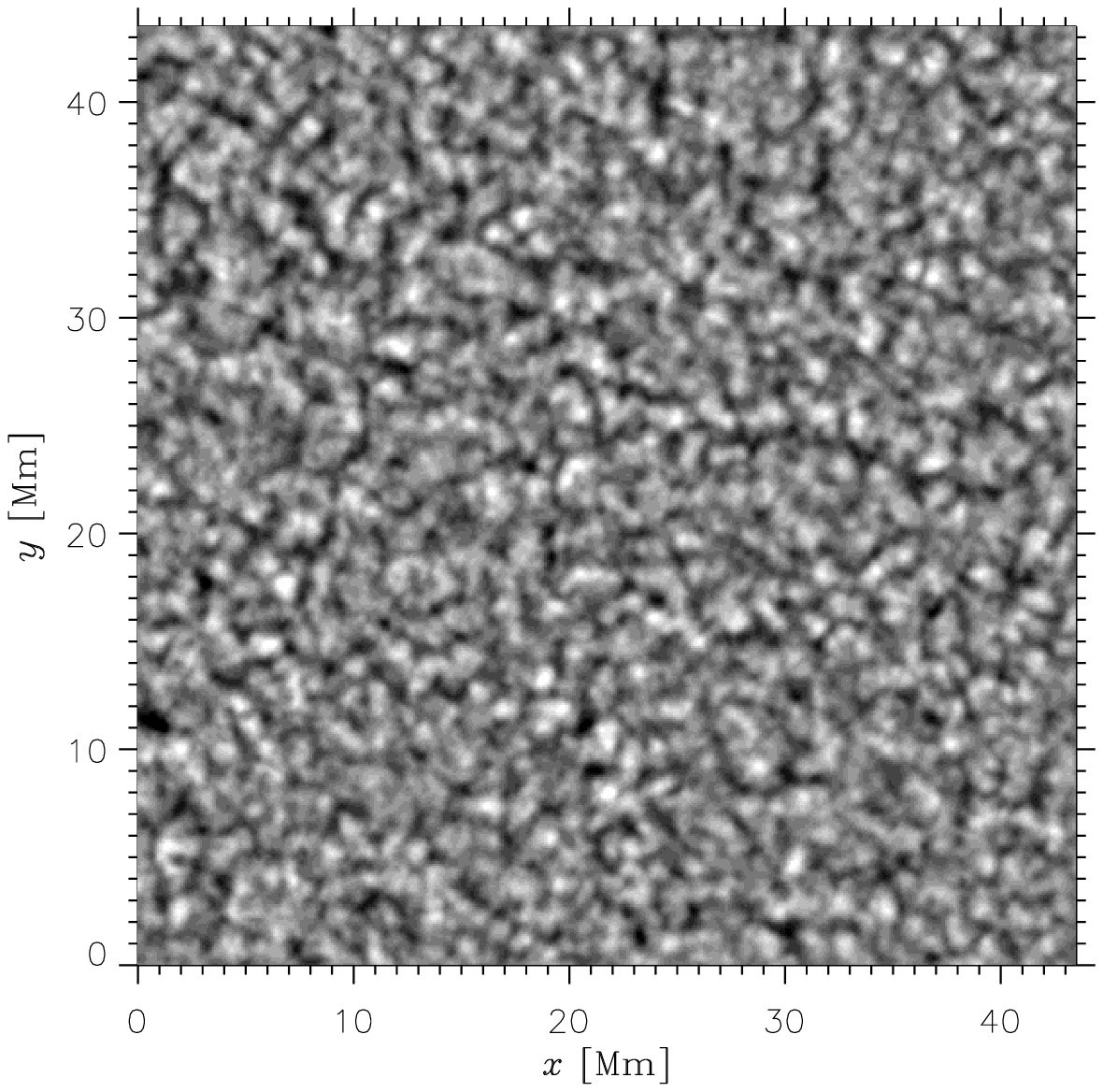}
 \includegraphics[width=0.48\textwidth]
 {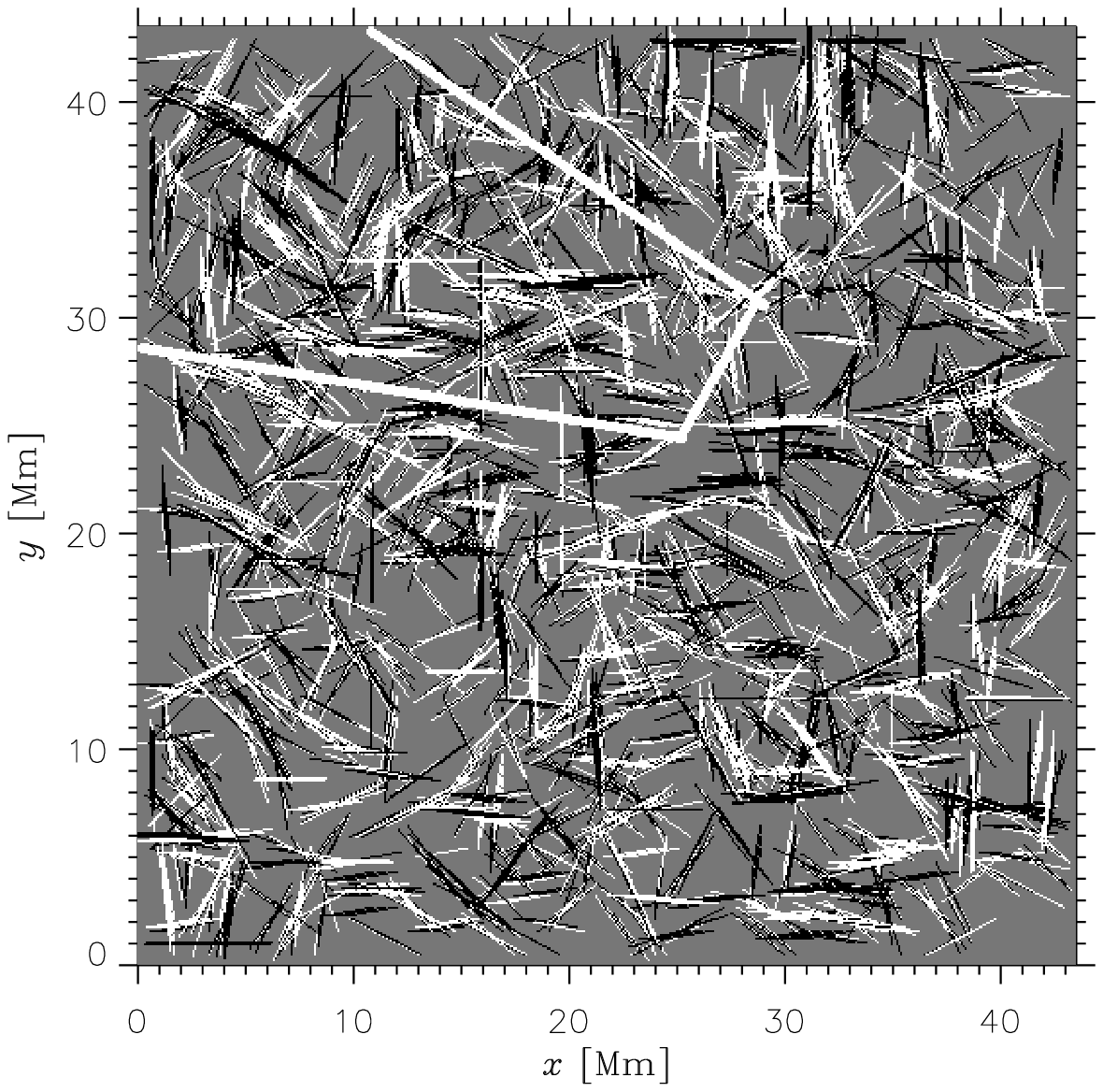}\\
 {\scriptsize \hspace{0.6cm} (c) \hspace{5.6cm} (d)}
 \includegraphics[width=0.48\textwidth]
 {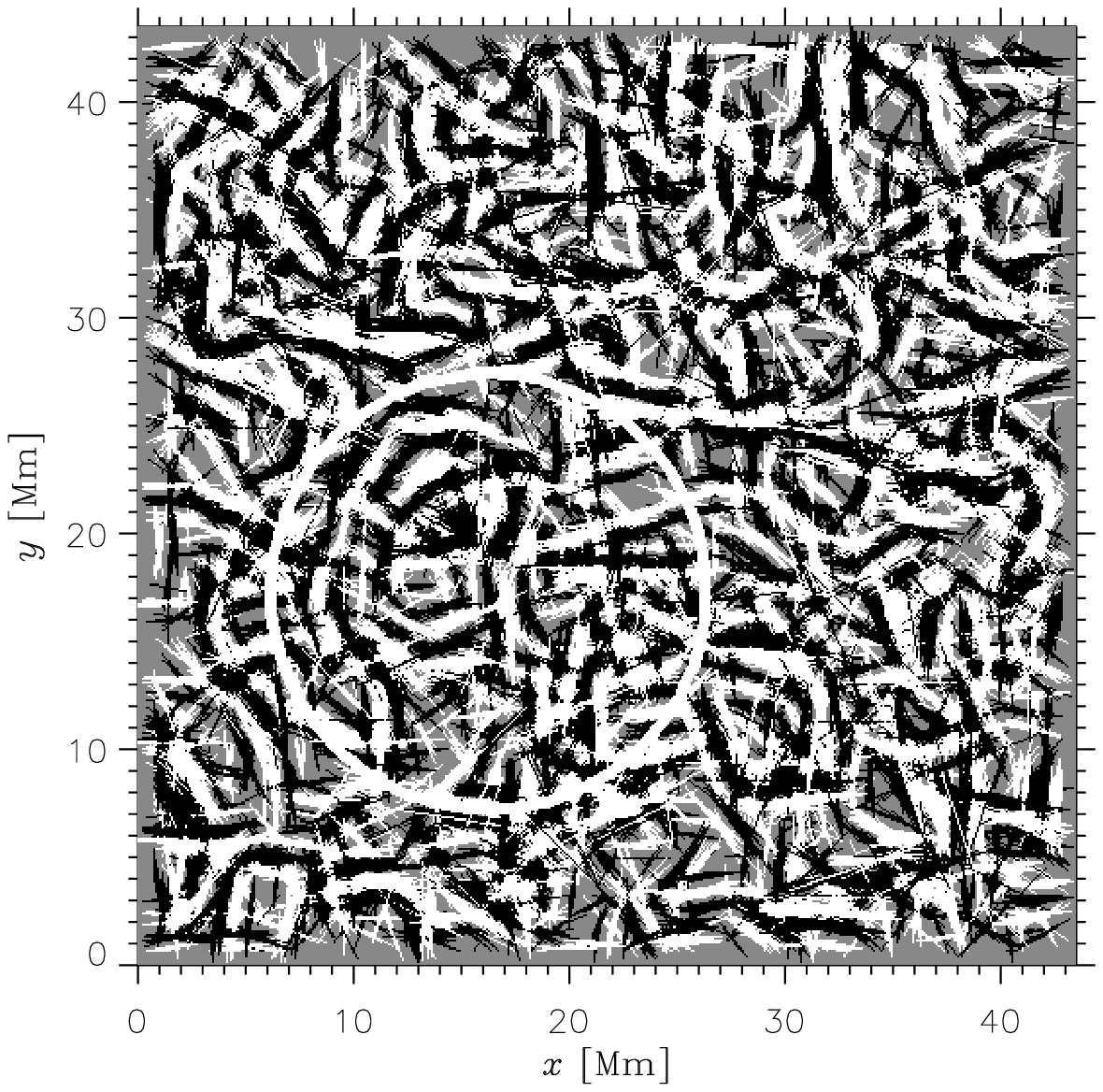}
 \includegraphics[width=0.48\textwidth]
 {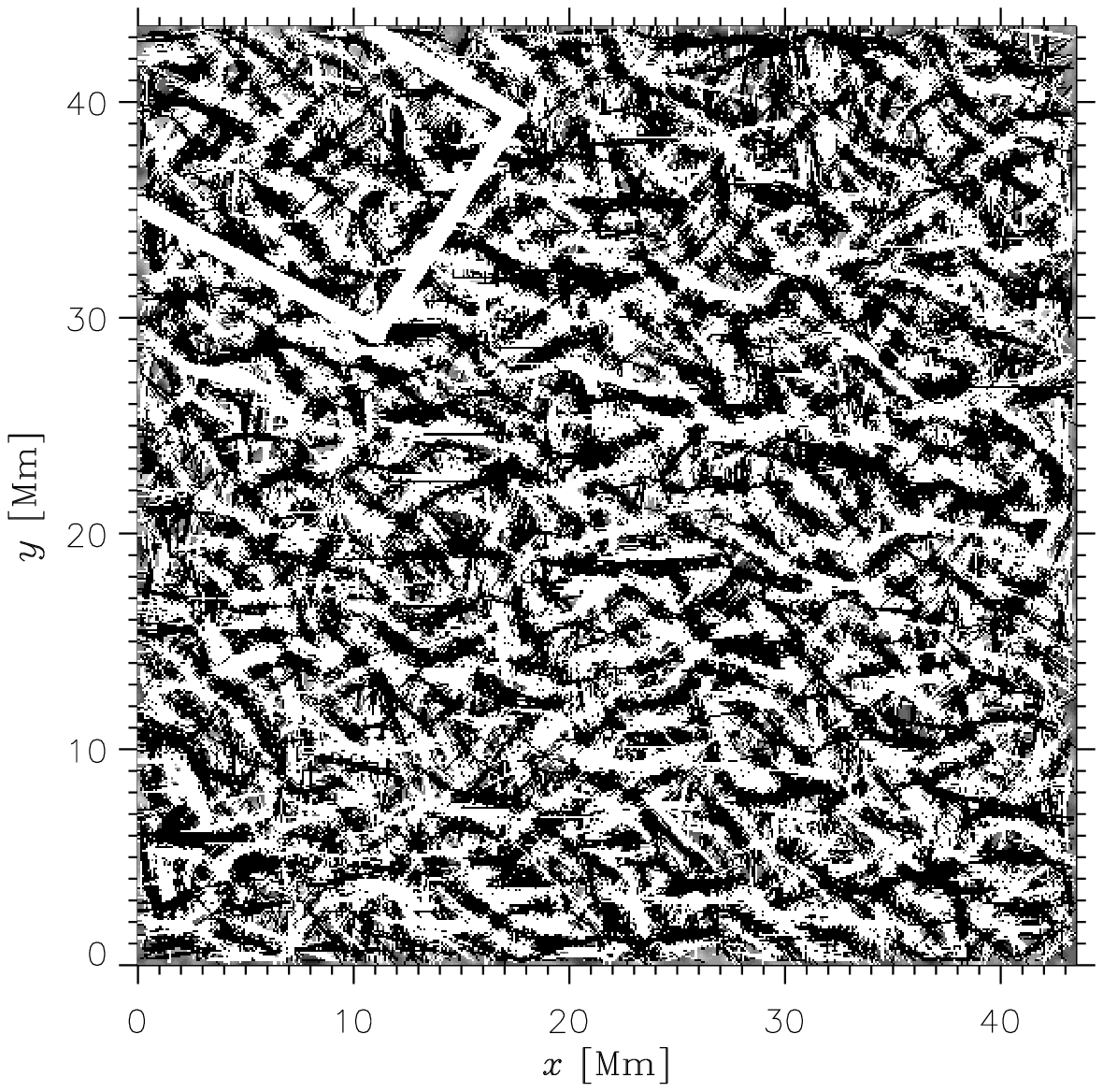}
  \caption{(a) A two-hour-averaged image and the results of contour
  detection in this image at (b) $p=2.5\times 10^{-4}$, $\Delta=5$,
  $l=16$\,--\,43; (c) $p=2\times 10^{-3}$, $\Delta=8$, $l=10$\,--\,30;
  and (d) $p=10^{-2}$, $\Delta=2$, $l=5$\,--\,20.}\label{fr65}
\end{figure}

\begin{figure} 
\centering
  \includegraphics[width=0.48\textwidth]
 {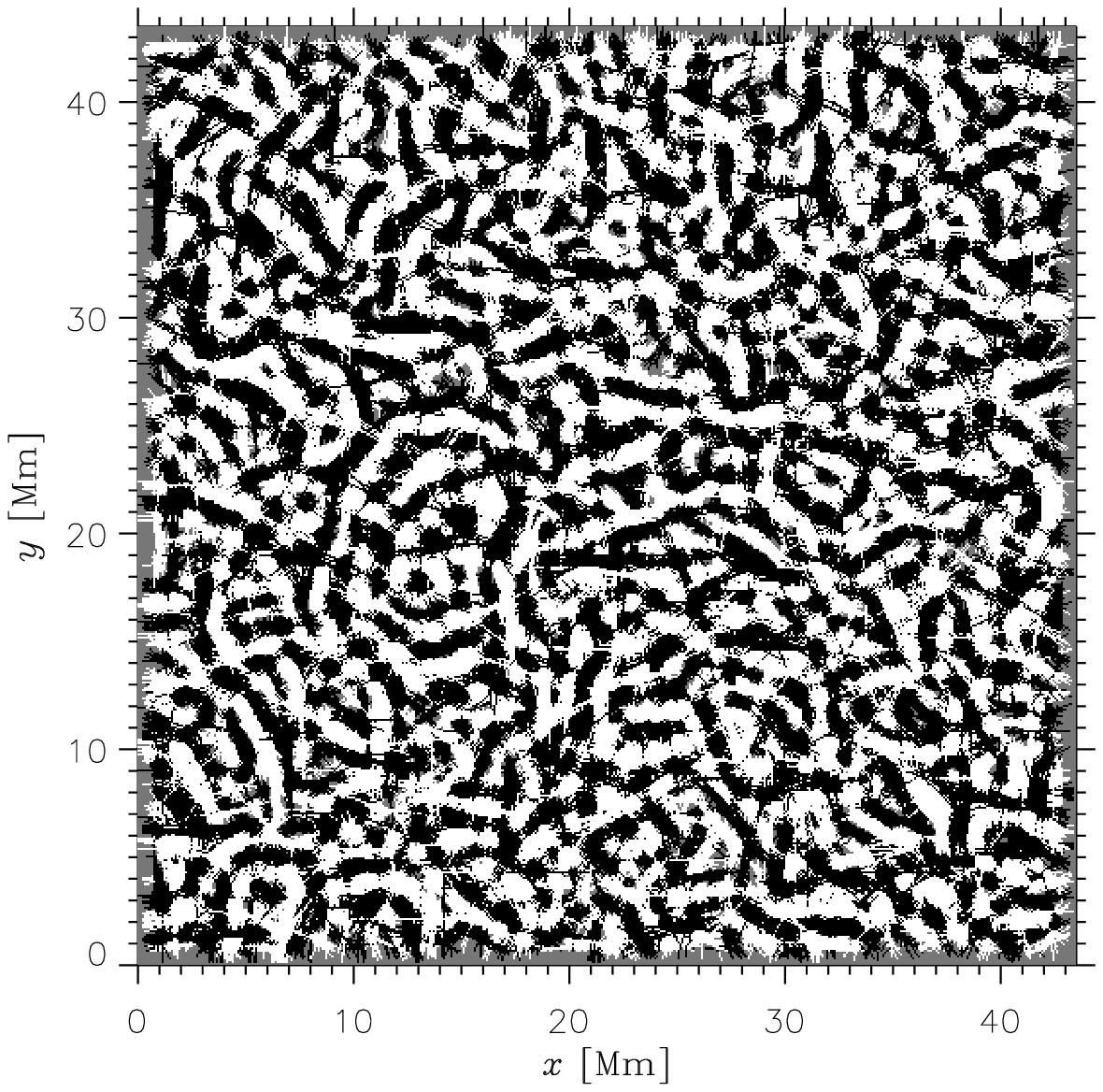}
 \caption{An illustration to the effect of the lineament-filtering procedure: the result of processing the image shown in Figure~\ref{fr65}a at the same values of the processing parameters as in Figure~\ref{fr65}c\protect\ but without filtering.}
 \label{fr65unfilt}
\end{figure}

First of all, the optimum $\Delta$ value depends on the width of
the linear features that are sought. Typically, we are
seeking contours that correspond to parallel ridges and trenches
in the relief of the brightness field. In the contour-detection
mode, the COLIBRI code draws contours along the slopes
and marks them as white or black curves depending on the direction
of the slope, \textit{i.e.,} on the stochastically-predominant sign of the
difference between the brightness values at the points located on
the two sides of the contour at the same distance $\Delta$.
Usually, the code draws bunches of nearly-parallel, closely-located contours of either sort, which merge into white or black
lanes. If $\Delta$ is small, the detected contours should gather
near the steepest-slope curve. However, as our experience shows
(and as confirmed by some qualitative considerations), at larger
$\Delta$ values comparable with some optimal, ``resonant''
distance related to the characteristic distance between ridges and
trenches, the algorithm can output white lanes highlighting the
ridges and dark lanes highlighting the trenches.

Similarly, there is no universally-preferable value of the
parameter $p$, which is regarded as the admissible probability of
spurious detection. The optimum $p$ is not universal and should be
properly chosen for any processing regime specified by the other
parameters.

The range of lineament lengths [$l$] used to construct contours
has also some effect on the detection of structures. Generally,
the algorithm can better integrate lineaments into contours for
broader $l$ ranges. However, the inclusion of very-short
lineaments results in a very high noise level. Moreover, the
computation time grows dramatically with widening the $l$ range.
Thus, a reasonable compromising choice should be made.

Let us consider the two-hour-averaged image shown in Figure~\ref{fr65}a
and the structures detected in this images
(Figures~\ref{fr65}b\,--\,\ref{fr65}d). The last three panels are
arranged in the order of increasing $p$; the largest $\Delta(=8)$
is used in the case of panel c and the smallest $\Delta(=2)$ in
the case of panel d. As we can see, numerous patches of
trenching patterns (not everywhere regular) with a characteristic
width of ridges comparable to the granular size are clearly
manifest in panel c. It is noteworthy that, in this figure, most
white lanes are paralleled by one or more other white neighbouring
lanes and at least two black neighbouring lanes. In other words,
trenching patterns are very common in the image (although they
vary widely in their area and in the number of alternating white
and black lanes). A careful inspection of panel c shows that the
black lanes really correspond to dark features in the images but
not merely to background areas where no features have been
detected (such areas are shown in grey). In panel d, the image
processing everywhere highlights finer details of the pattern. As
a rule, they are even less regular than the granular-scaled
ridges; however, a system of very narrow parallel ridges can be
distinguished near the upper left corner of the frame, within the
(incomplete) white rectangle, as a pattern of very thin hatches
inclined to the left. Thus, as we would expect, the characteristic
scale of detectable features decreases with the decrease of
$\Delta$. On the other hand, if we reduce $p$ by an order of
magnitude (see panel b for example), too many features
disappear, being treated by the code as ``spurious''. The
remaining segments of curves are only the ``most reliably
identified'' fragments of the structures, most parts of which have
been filtered out. Nevertheless, even such drastically filtered
patterns may be useful: the pattern enclosed in the (incomplete)
white trapezium in the upper left part of panel b suggests that
a combination of concentric arcs and radial rays is present there,
and it can be classified as a specific form of web pattern (the
occurrence of such patterns in averaged images was noted by
Getling, 2006).

\begin{figure} 
\centering
 {\scriptsize \hspace{0.6cm} (a) \hspace{5.6cm} (b)}
  \includegraphics[width=0.48\textwidth]
 {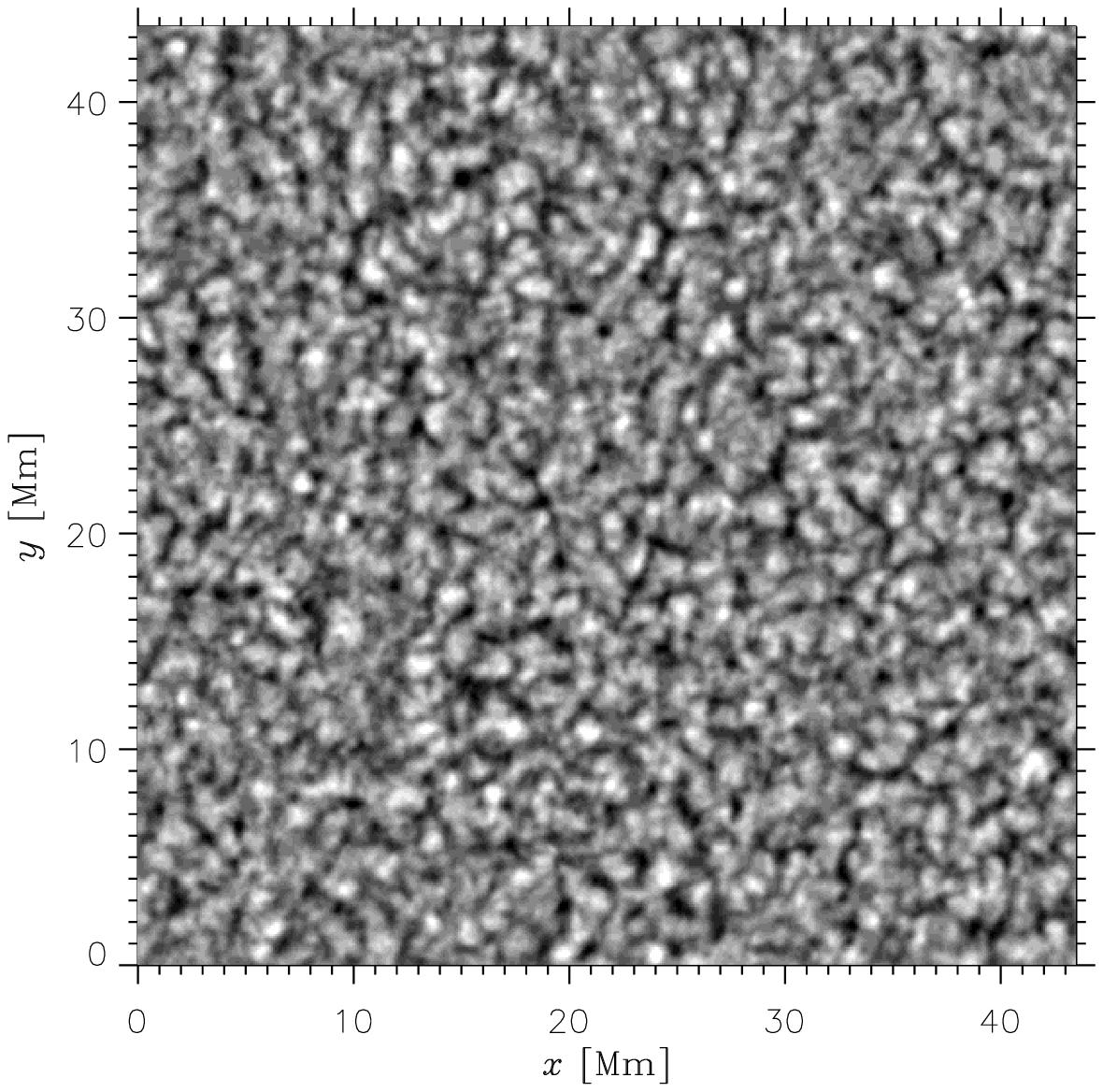}
 \includegraphics[width=0.48\textwidth]
 {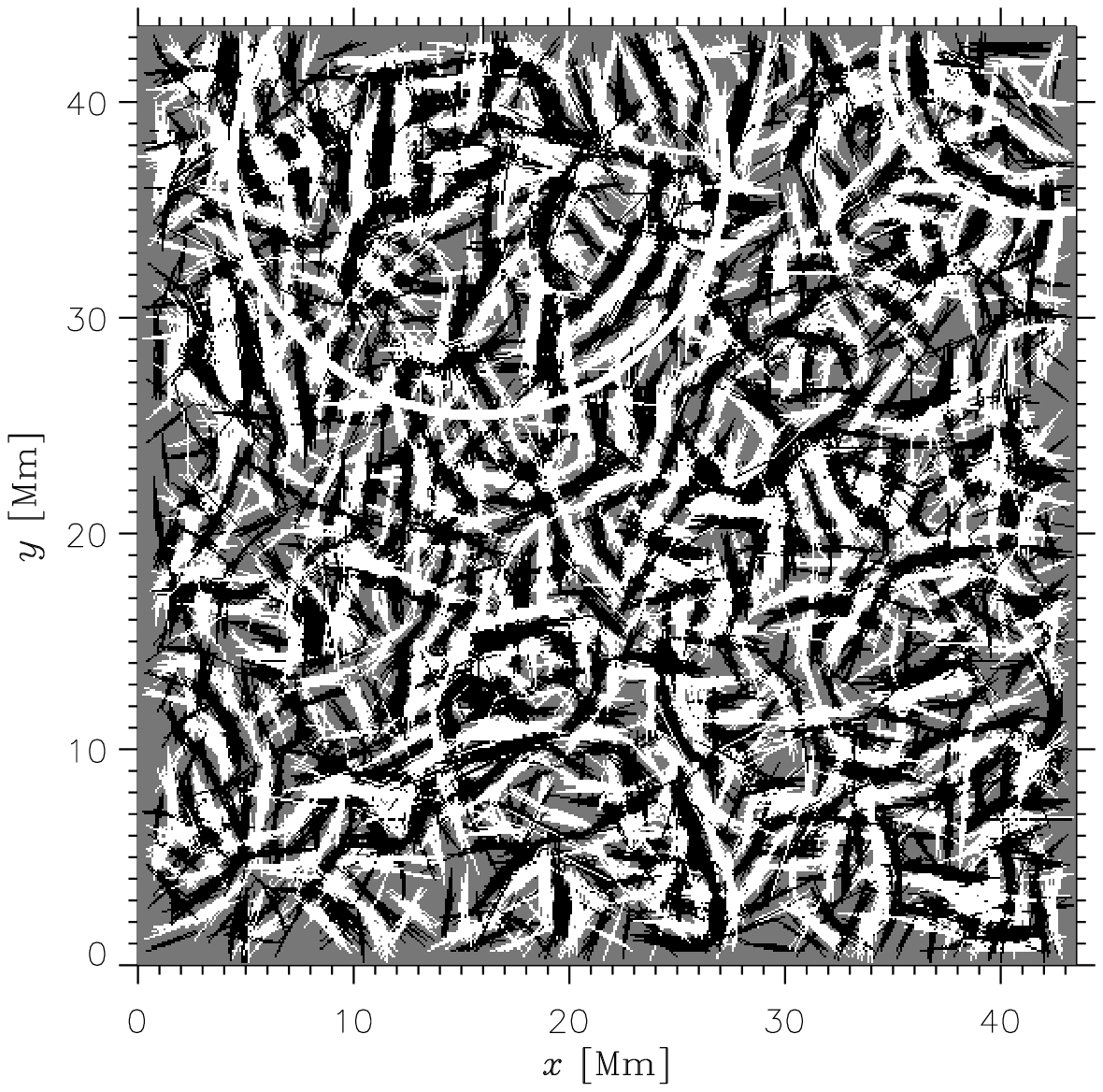}\\
 {\scriptsize \hspace{0.6cm} (c) \hspace{5.6cm} (d)}
 \includegraphics[width=0.48\textwidth]
 {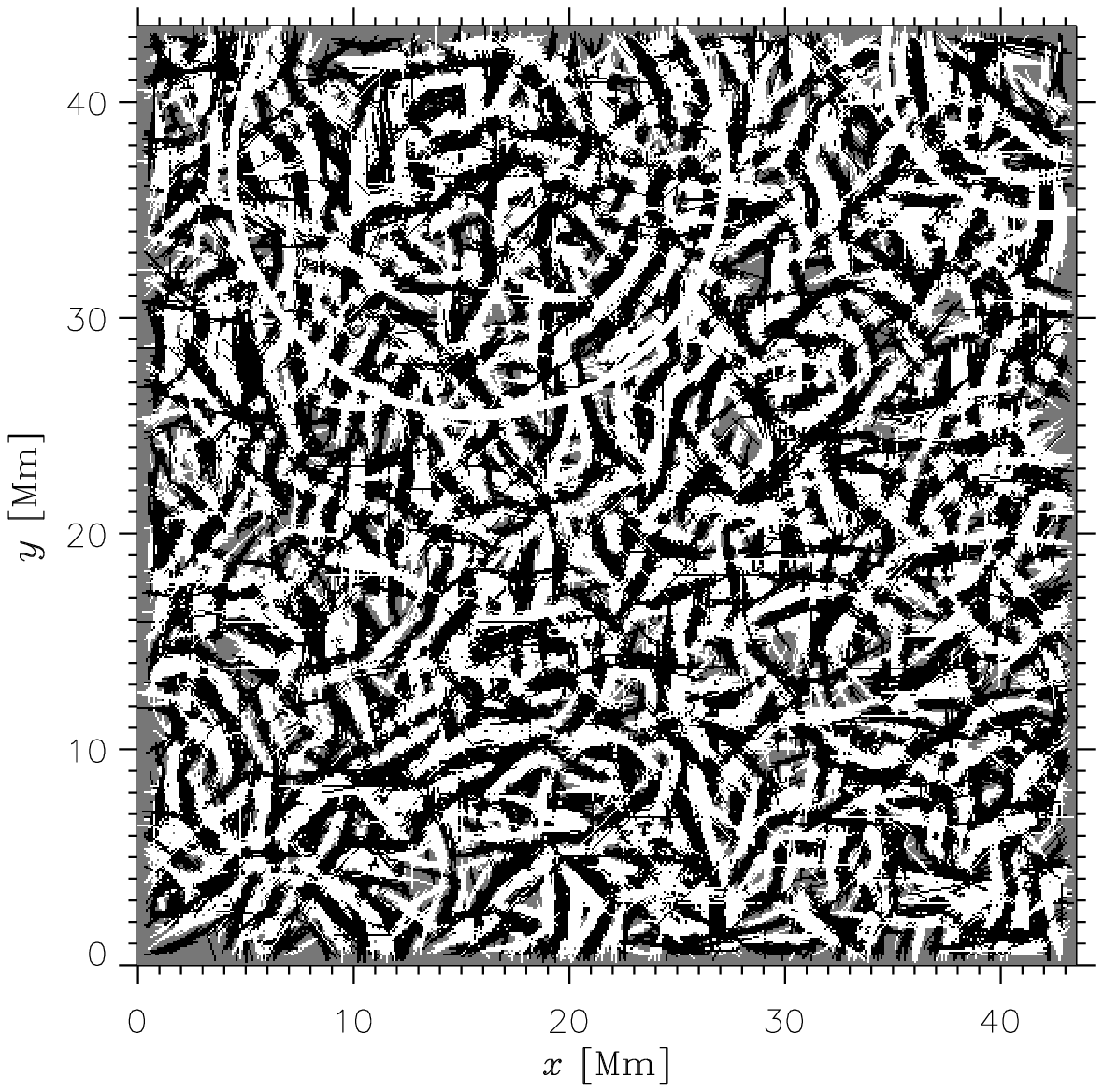}
 \includegraphics[width=0.48\textwidth]
 {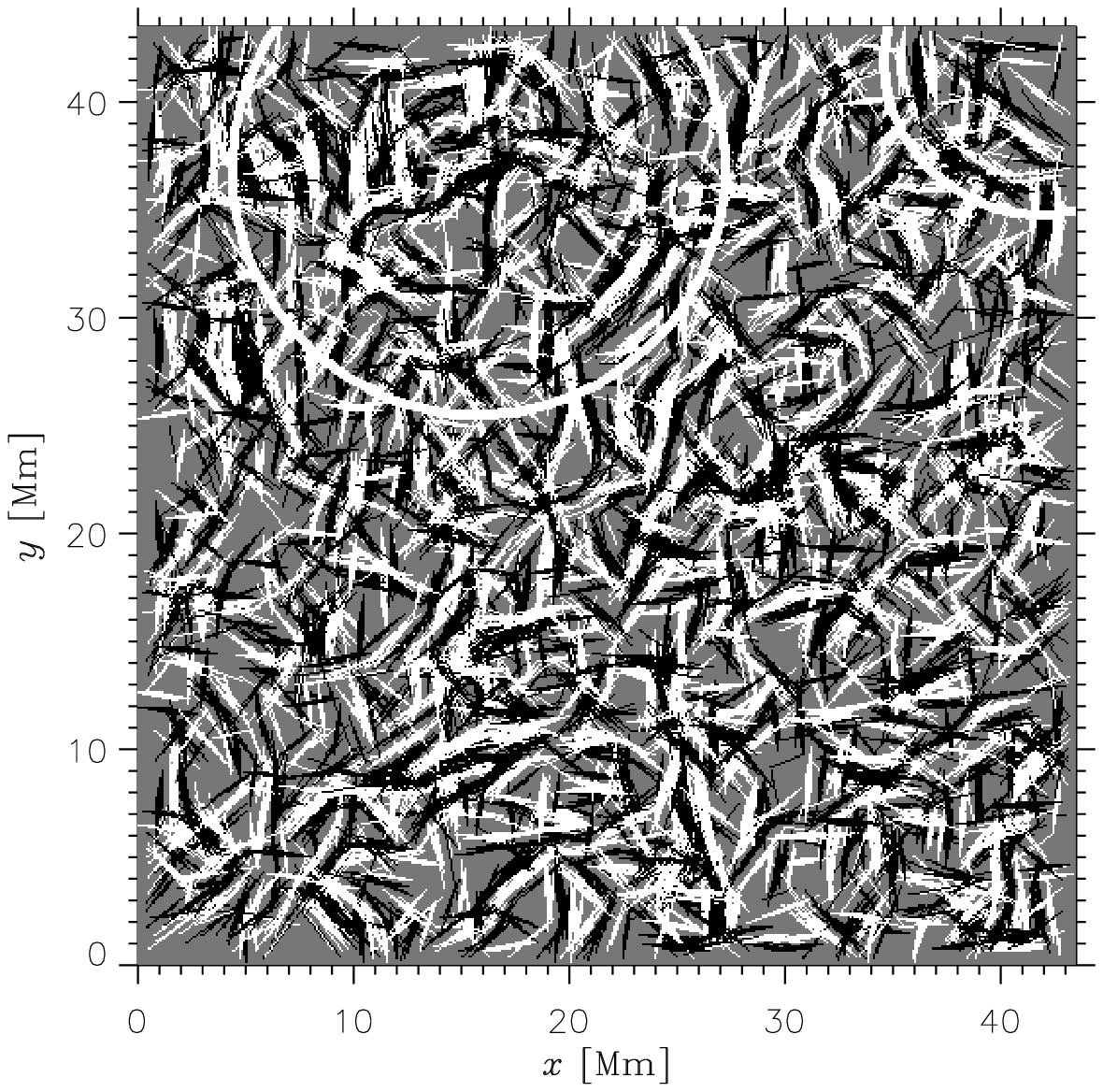}
  \caption{(a) Another two-hour-averaged image and the results of
  contour detection in it at (b) $p=2\times 10^{-3}$, $\Delta=8$,
  $l=10$\,--\,30; (c) $p=10^{-2}$, $\Delta=6$, $l=10$\,--\,20,
  and (d) $p=10^{-2}$, $\Delta=6$, $l=20$.}\label{fr107}
\end{figure}

In the pattern of granular-scaled trenching, which is most
pronounced in panel c, the structure marked by the white circle
deserves particular attention. It appears as a system of
concentric rings deformed and ``spoiled'' in some way and should
therefore be included in our collection of ring systems. At the
same time, it provides an example of a structure that can hardly
be visually distinguished in the original image but which becomes
detectable if the image is processed with a properly chosen
parameter $\Delta(=8)$.

To illustrate the effect of the lineament-filtering procedure, we present here in Figure~\ref{fr65unfilt} the result of processing the image shown in Figure~\ref{fr65}a at the parameters used to obtain Figure~\ref{fr65}c but without filtering. It can easily be seen that the filtering procedure efficiently removes isolated bright blotches and very short lineaments, preserving more extended features and areas with pronounced trenching.

\begin{figure} 
\centering
 {\scriptsize \hspace{0.6cm} (a) \hspace{5.6cm} (b)}
 \includegraphics[width=0.48\textwidth]
 {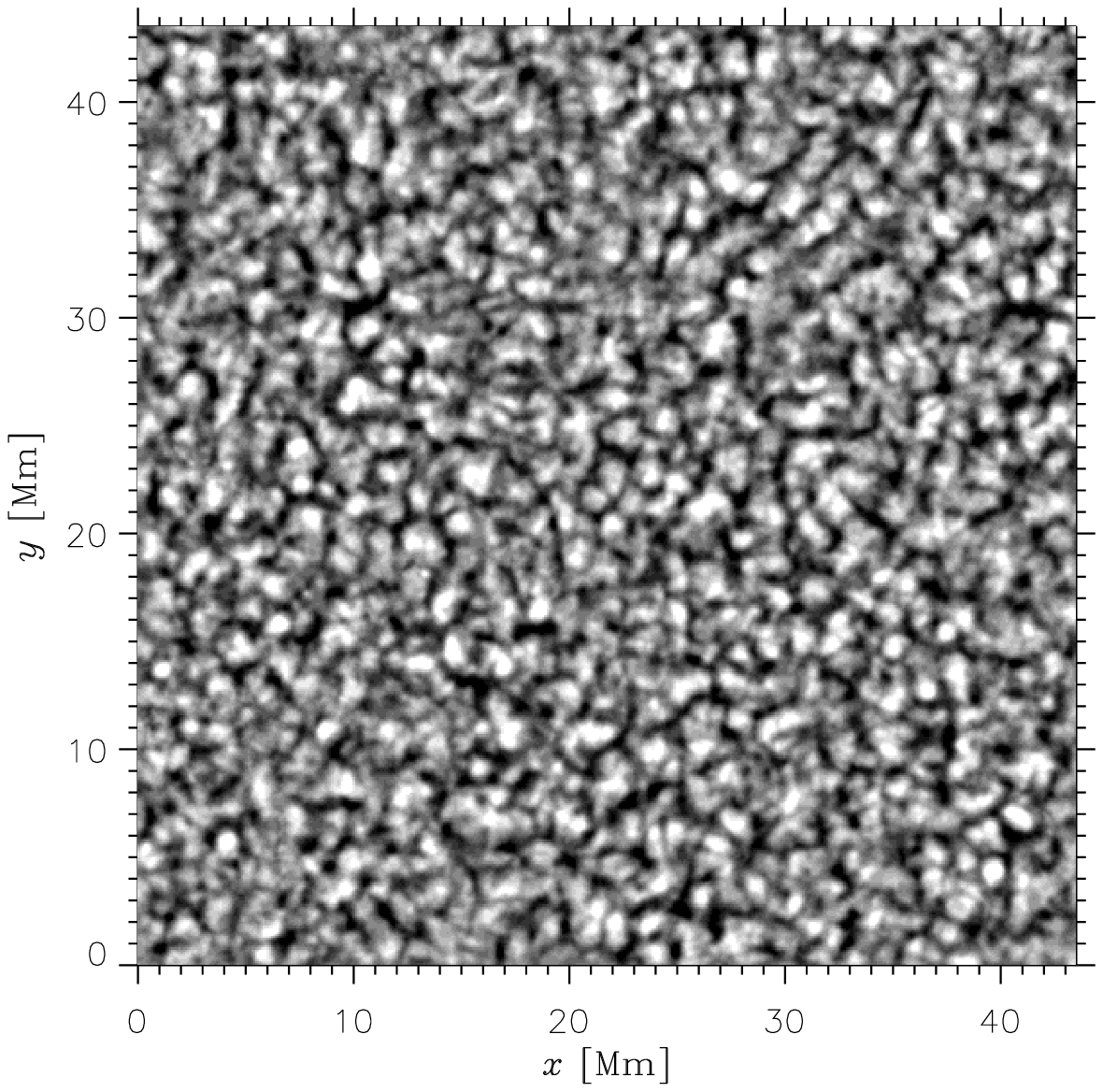}
 \includegraphics[width=0.48\textwidth]
 {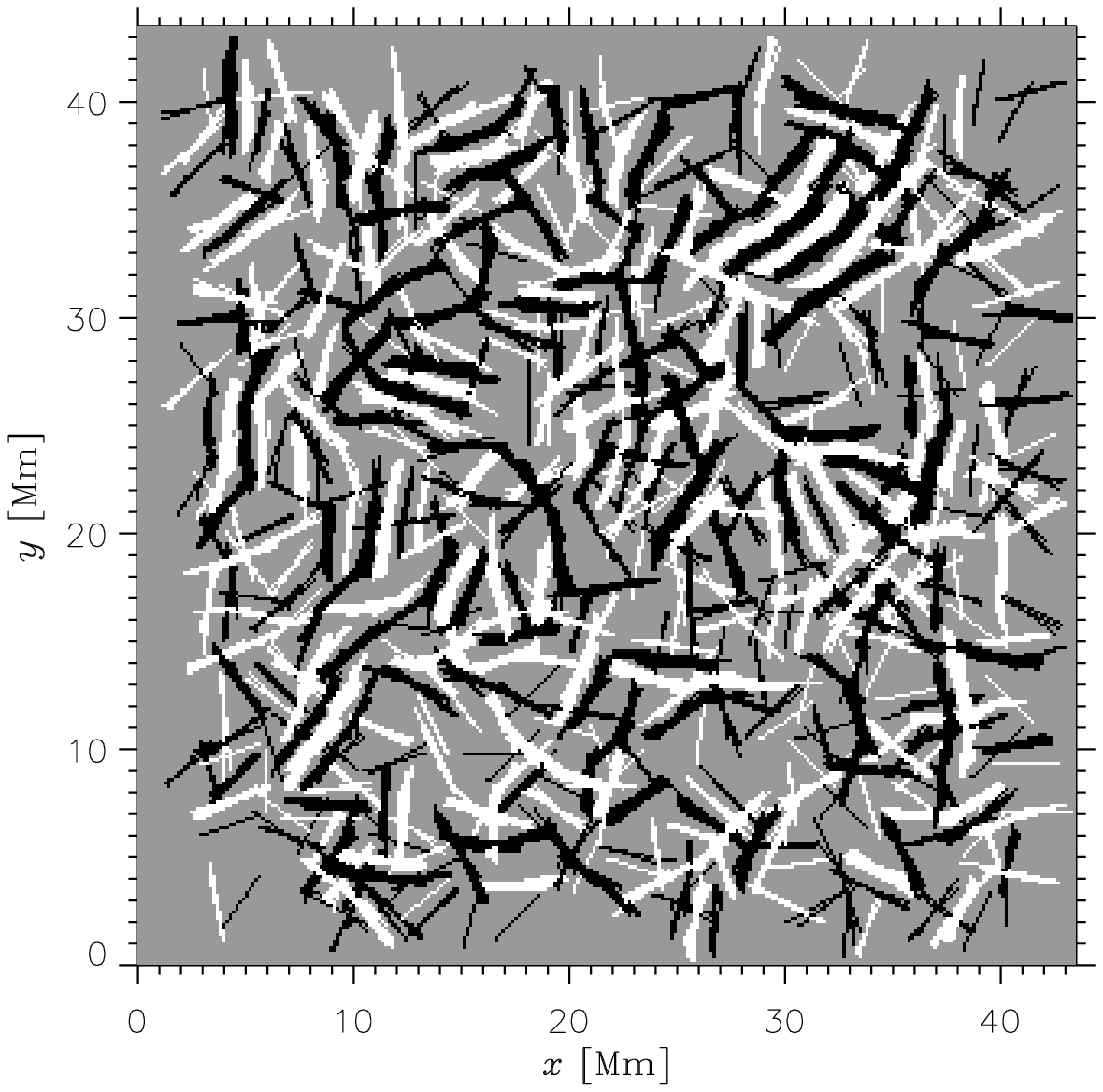}\\
 {\scriptsize \hspace{0.6cm} (c) \hspace{5.6cm} (d)}
 \includegraphics[width=0.48\textwidth]
 {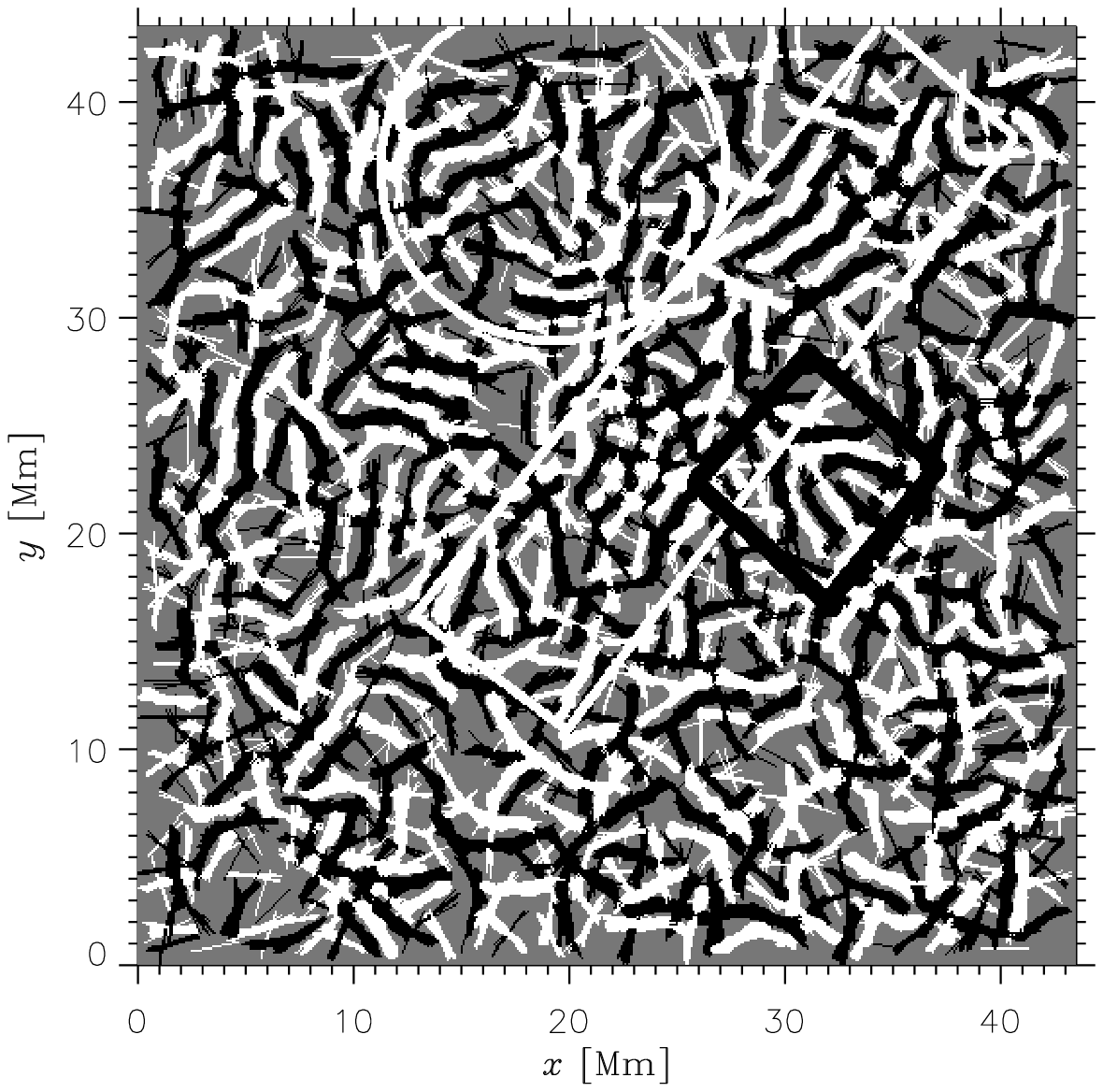}
 \includegraphics[width=0.48\textwidth]
 {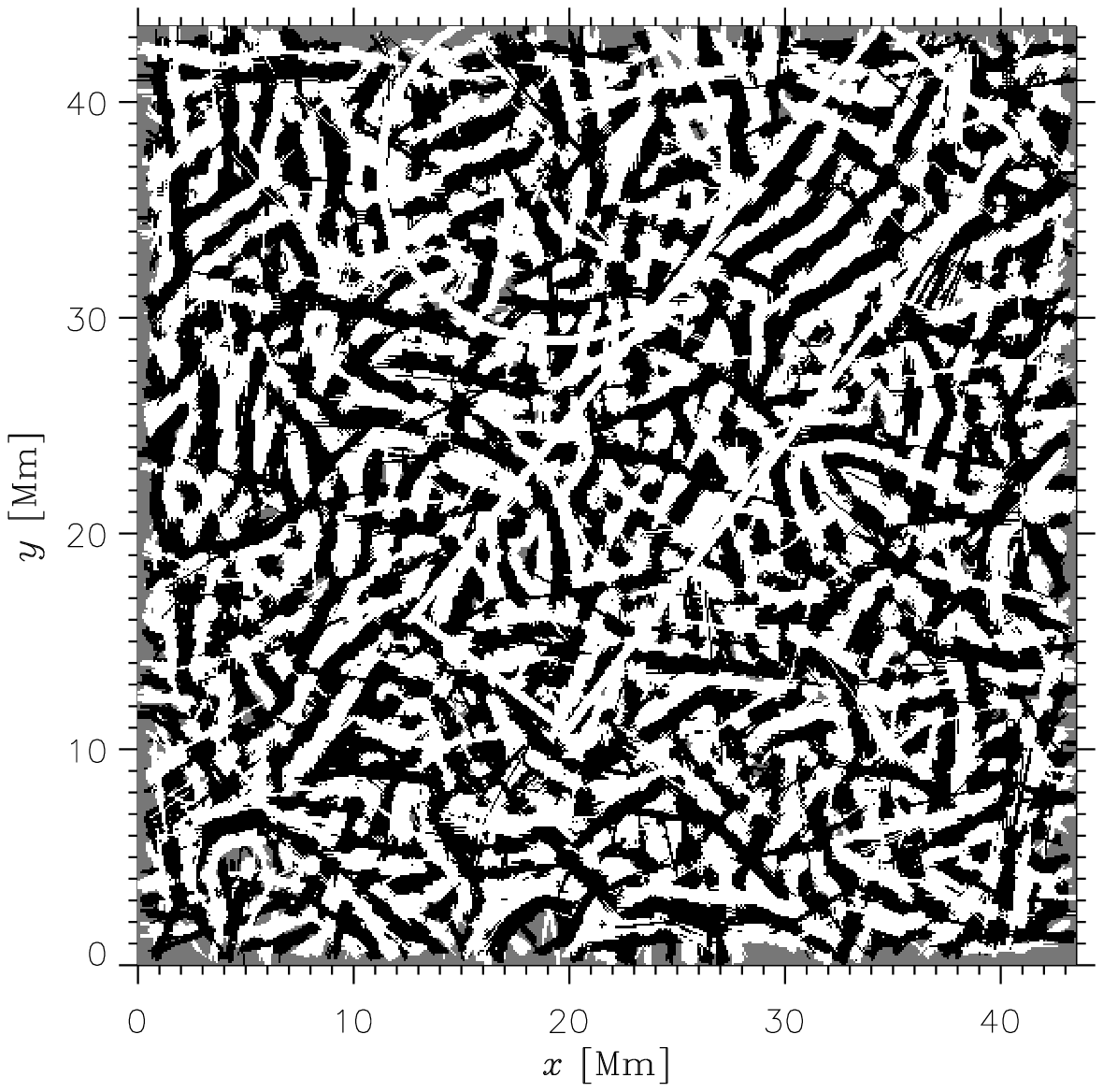}
  \caption{An image averaged over a one-hour interval centred at nearly
  the same time as in the case of Figure \ref{fr65}\protect\ and
  a comparison between the regimes of lineament and contour
  detection: (a) original image; (b) lineaments detected at
  $p=10^{-3}$, $\Delta=5$, $l=16$\,--\,43; (c) lineaments detected at
  $p=2.5\times 10^{-4}$, $\Delta=8$, $l=16$\,--\,43; (d) contours detected
  at $p=10^{-4}$, $\Delta=8$, $l=16$\,--\,43.}
  \label{fr104}
\end{figure}

\begin{figure} 
\centering
 \includegraphics[width=0.4\textwidth]
 {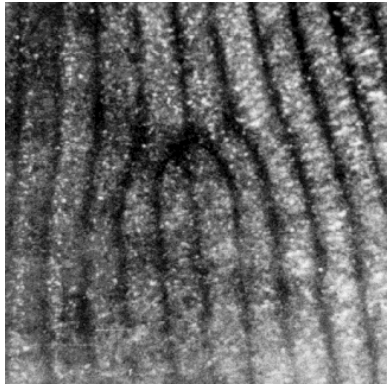}
 \caption{Dislocation in a roll-convection pattern (experimental
 photograph by Berdnikov and Markov).}
 \label{disloc}
\end{figure}

Figure \ref{fr107} also refers to two-hour averaging. Here, panel a
shows an averaged image for which the midtime of the
averaging interval is nearly the same as for Figure~3 in Getling and Brandt
\cite{gb1}. This is still the best-available (most interesting)
image, in terms of the presence of pronounced and highly ordered
structures. The pattern in panel b was obtained using the same
parameters as in the case of Figure~\ref{fr65}c. Encircled in the
upper left quadrant is the well-developed system of concentric
rings that appear as fairly regular circles in panel a.
Because of the presence of dark radial gaps interrupting the
circles, the algorithm in some cases combines fragments of
different rings into single contours, distorting the visual
regularity of the system. A structure that may be part of another
ring system, also marked with an (incomplete) circle, can be seen
near the upper right corner of the frame. As in
Figure~\ref{fr65}c, even outside the regular structures, white
lanes are accompanied by and alternate with black ones, and vice
versa. In panel c, finer details are highlighted, since a
smaller $\Delta$ is used in this case; in addition, $p$ is here
higher than for panel b. It is important that the main features
present in the pattern do not disappear as $p$ is reduced from
$10^{-2}$ (c) to $2\times 10^{-3}$ (b), although the simultaneous
increase of $\Delta$ additionally removes the narrowest lanes.

Finally, the effect of the $l$ range can be grasped by comparing
panels c and d. The patterns shown in these panels were
obtained at the same $p$ and $\Delta$ but the range $l=10$\,--\,20 was
shrunk to the single value $l=20$ with passing from panel c to panel d.
Accordingly, short lineaments disappeared; however, objects
exceeding 20 pixels in lengths were preserved because of combining
partially overlapping lineaments into longer ones in the process
of filtering and combining chains of lineaments into contours.

Thus, different combinations of parameter values can highlight
different features in the image. To form a more complete idea of
the structures present in the granulation field, the parameters at
which the image is processed should be varied.

Whereas varying the image-processing parameters can be useful for
the detection of features differing in their size, varying the
averaging time makes it possible to reveal features differing in
their lifetime. In this respect, it appears instructive to compare
Figure \ref{fr107}, for which the averaging time is two hours, with
Figure \ref{fr104}, for which this time is one hour and the averaging
interval is centred at virtually the same time as in the case of
Figure \ref{fr107}.\footnote{In Figure \ref{fr104}, where results
obtained in both the lineament-detection and contour-detection
modes are presented, they, by chance, appear very similar. In no
way is this the general situation.}

The original averaged image is given in Figure \ref{fr104}a; panels b and
c present the results of processing it in a lineament-detection
mode without filtering.\footnote{The pattern shown in Figure
\ref{fr104}b was obtained using an older version of the code,
which could not draw lineaments near the edges of the field of
view.} In both panels, there are many patches with a highly
pronounced trenching. Panel c is considerably richer in
ridge--trench systems than panel b, although the corresponding
$p$ value is four times smaller than for panel b. This is an
additional illustration of the role of the parameter $\Delta$. As
in the above-considered examples, the value $\Delta=8$ is most
favourable for the detection of contours of the fundamental width,
and this fact proves here to be more important than some reduction
of $p$.

\begin{figure} 
\centering
 {\scriptsize \hspace{0.6cm} (a) \hspace{5.6cm} (b)}\\
 \includegraphics[width=0.48\textwidth]
 {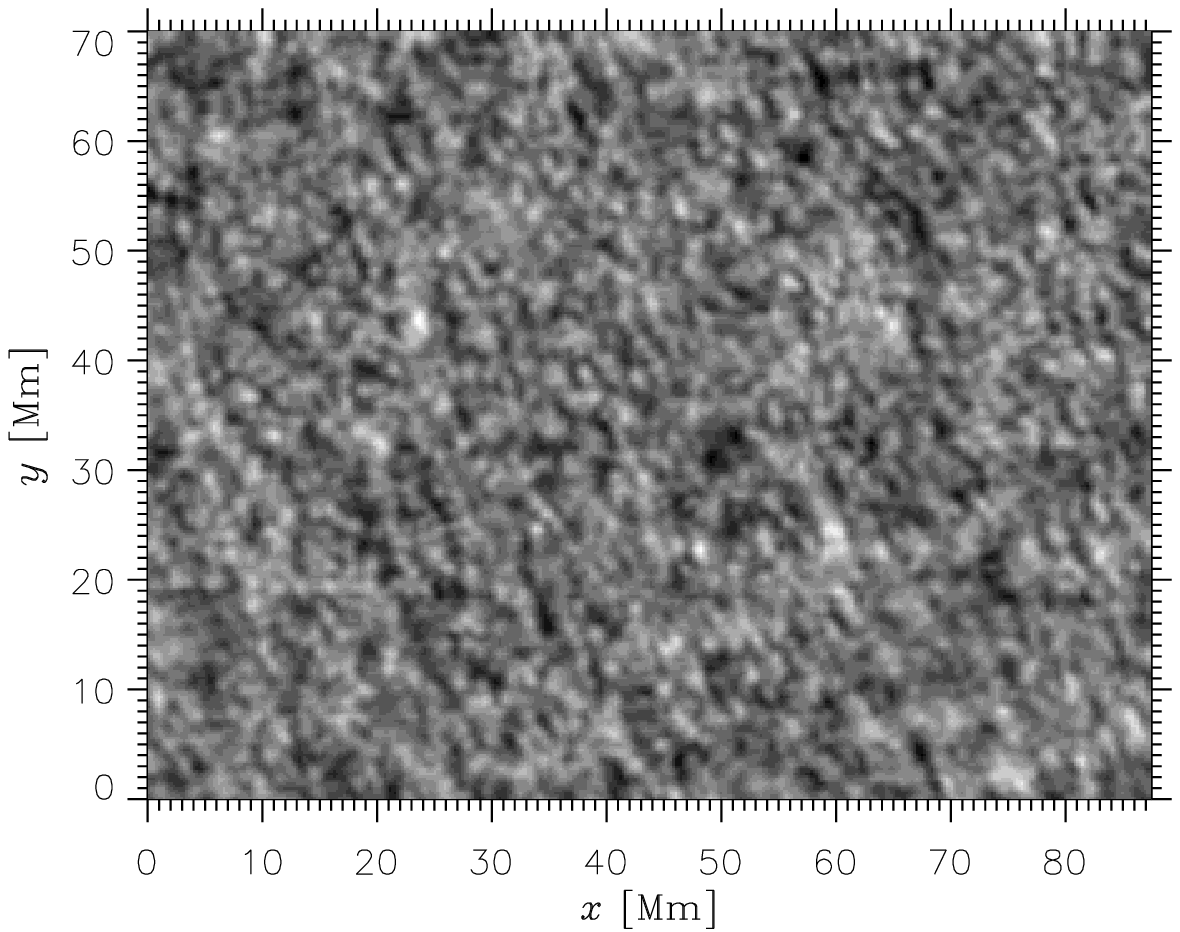}
 \includegraphics[width=0.48\textwidth]
 {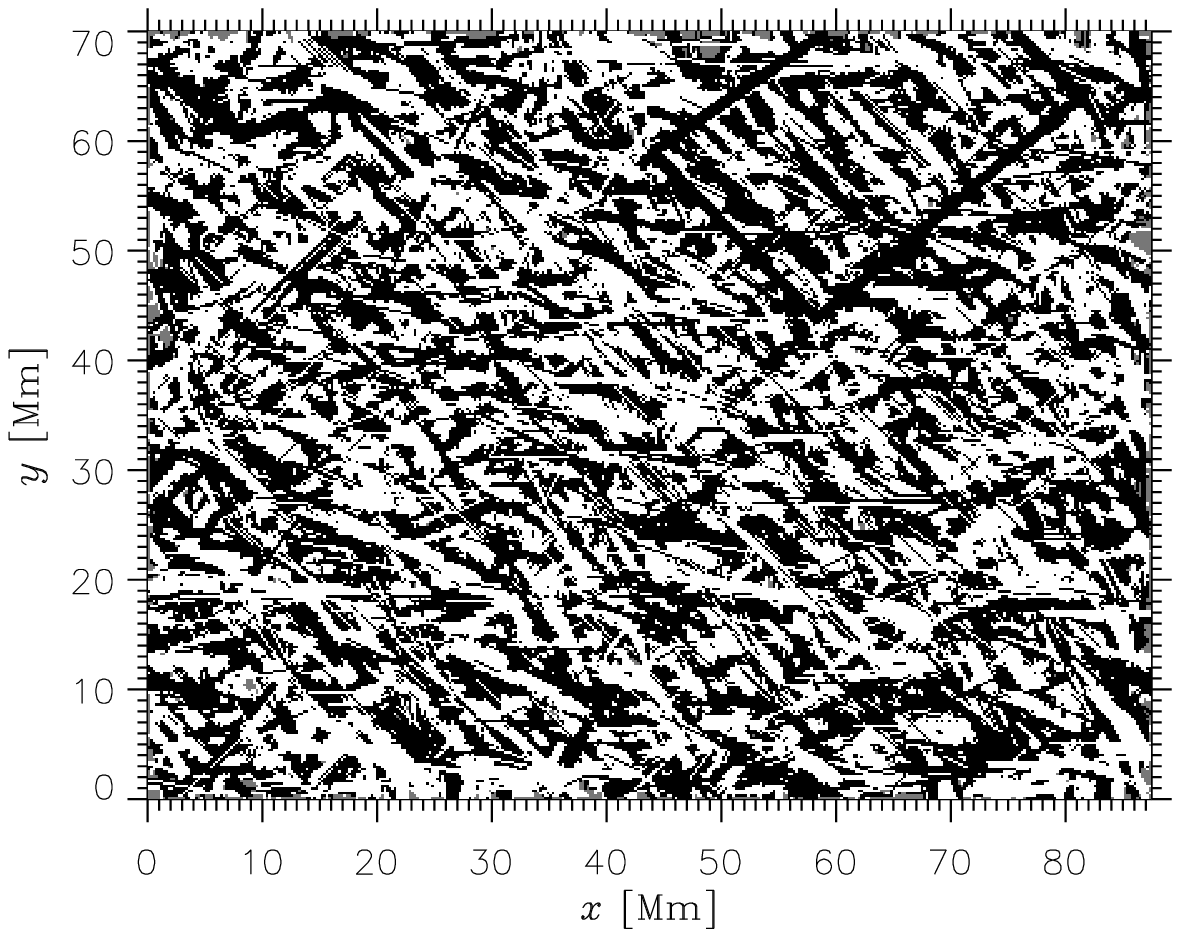}
 \caption{Processing of a two-hour-averaged MDI image: (a) original image;
 (b) the results of contour detection at $p=5\times 10^{-3}$,
 $\Delta=2$, $l=20$.}
 \label{MDI}
\end{figure}

It is interesting that the structure marked with a black box in
Figure~\ref{fr104}c is very similar to a fragment of a
convection-roll pattern with a dislocation. Such patterns have
received much attention in studies (particularly, experimental) of
thermal convection (see, \textit{e.g.,} Getling, 1998, for a survey); for
comparison, we present an experimental photograph of a
roll-convection pattern with a dislocation in Figure \ref{disloc}. (Obviously, this is merely an illustrative example; taken alone, it is insufficient to claim that the observed feature is actually related to roll convection.)

The white rectangular box in Figures \ref{fr104}c and \ref{fr104}d
(the latter panel representing the results of contour detection)
marks an extended ridge--trench system, which is almost completely
smeared in the two-hour-averaged image (Figure \ref{fr107}). The
pattern within the white circle bears only faint resemblance to
the pattern marked with the circle in Figures
\ref{fr107}b\,--\,\ref{fr107}d, although some circular arcs can
apparently be associated with the well-developed system present in
Figure \ref{fr107}.

These observations clearly demonstrate that different features can
be identified in the averaged images depending on the length of
the averaging interval. In the case at hand, the system of
concentric rings visible in Figure~\ref{fr107} appears to have a
longer lifetime than has the elongated system identifiable in Figure
\ref{fr104}.

Finally, we give here an example of processing results for an
averaged image of the SOHO/MDI series (Figure \ref{MDI}). Panel
a is an enlarged cutout of the subsonically filtered images of
the MDI series averaged over a two-hour interval. As already
noted by Getling \cite{g06}, this averaged image (also reproduced
in that paper) exhibits a pattern of ridges and trenches in the
form of overall ``hatching'' inclined to the right. The pixel size
in the MDI series is about five times as large as in the La Palma
series; accordingly, $\Delta=2$ proves to be the optimum value for
the detection of contours. The contour-detection results obtained
at this value are shown in panel b. Generally, the 1.2$\arcsec$
MDI resolution is insufficient for our algorithmic processing, and
the code fails to detect the concentric-ring structures visible in
panel a (and marked in Figure 1 in Getling, 2006). However, in
some areas of the image trenching patterns can be detected with
certainty. A pattern of parallel ridges and trenches is
highlighted most distinctly within the black box in the upper
right quadrant of the field of view, where it appears highly
ordered.

\begin{figure} 
\centering
 {\scriptsize (a) \hspace{3.5cm} (b) \hspace{3.5cm} (c)}\\[2pt]
 \includegraphics[width=0.32\textwidth,bb=6 0 360 360pt,clip]
 {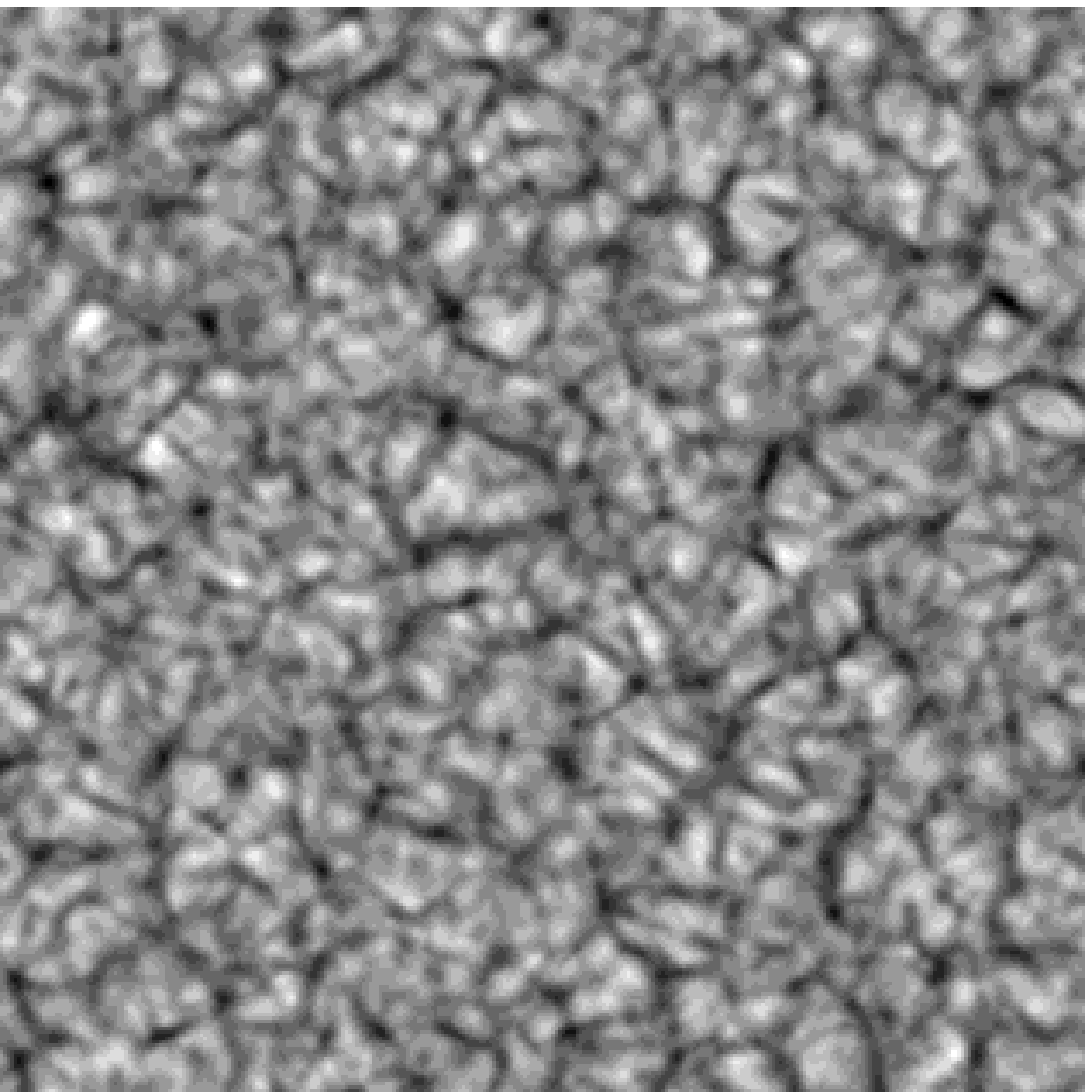}
 \includegraphics[width=0.32\textwidth,bb=6 0 360 360pt,clip]
 {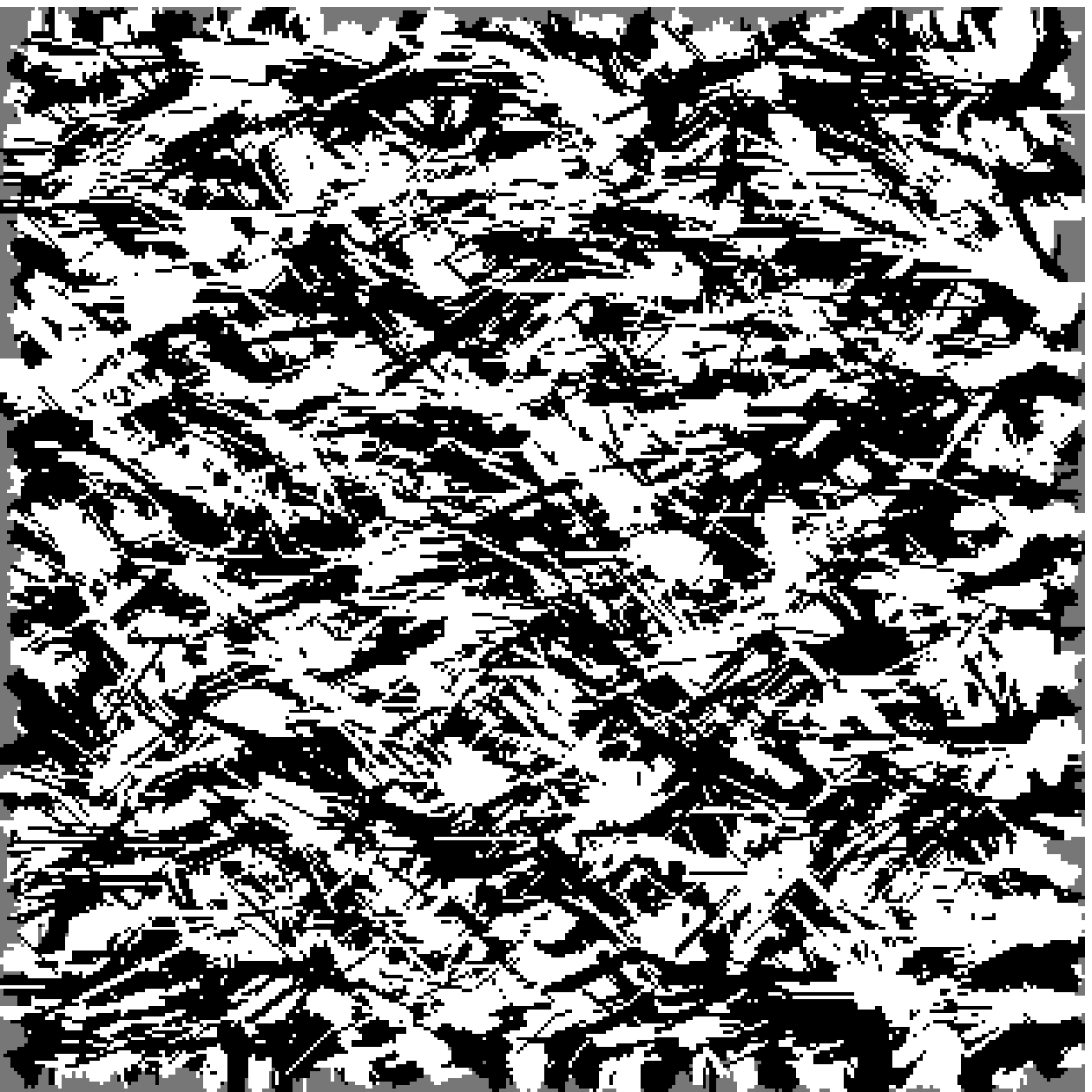}
 \includegraphics[width=0.32\textwidth,bb=6 0 360 360pt,clip]
 {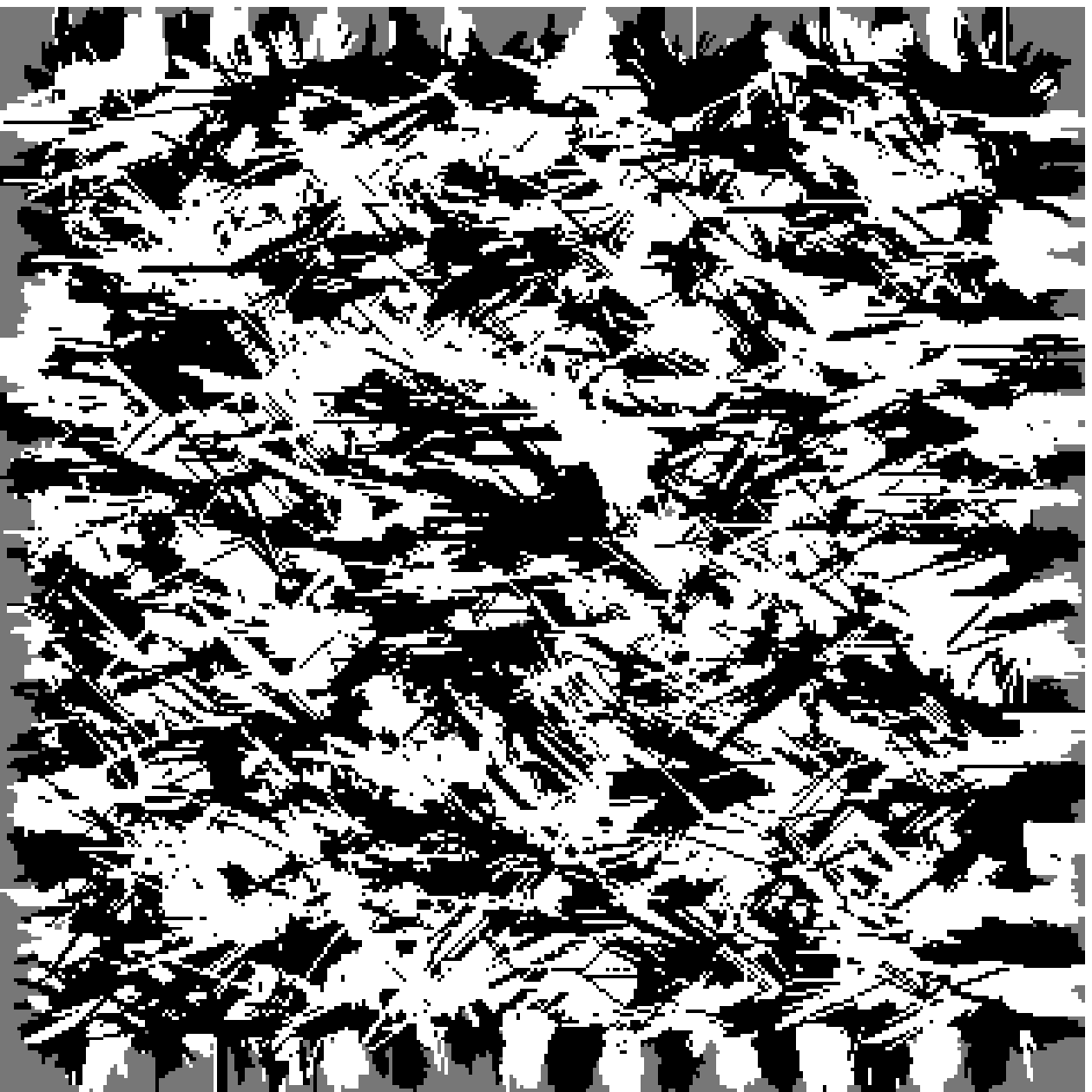}
 \caption{Processing of a simulated granulation field: (a) a two-hour average of the pattern obtained in simulations by Rieutord \etal \ (2002) and the results of contour detection in this averaged image with filtering at (b) $p=0.01$, $\Delta=8$, $l=20$ and (c) at $p=0.002$\,--\,0.01, $\Delta=16$, $l=20$.}
 \label{rieutproc}
\end{figure}

\section{Conclusion}

We have seen that the processing of time-averaged granulation
images by the COLIBRI code is capable of detecting
systems of ridges and trenches in the relief of the brightness
field, which vary in their characteristic scale, geometry, and
topology. The white lanes that highlight ridges are as a rule paralleled by one or more other white
neighbouring lanes and at least two black lanes that mark
trenches. Thus, our most general and remarkable finding is the
fact that trenching patterns, irrespective of their particular
geometries, are virtually ubiquitous in the averaged granulation
images. The patterns may be more or less regular, a patchy
appearance is typical of them, but the property that they include
alternating parallel white and black contours appears to be
universal (provided the images analysed by us are representative).

The detection of objects differing in their sizes and other
properties is possible if the parameters of image processing are
properly chosen. Of particular importance is the adjustment of the
parameter $\Delta$ to the characteristic width of ridges and
trenches.

The patterns analysed here can most naturally be interpreted in
terms of the structure of convective flows in the subphotospheric
layers. If this interpretation is correct, the light ridges in the brightness field should correspond
to material upflows, while the dark trenches can be associated
with downflows. Accordingly, the parallelism of alternating ridges
and trenches should be typical of roll convection, and, in the framework of our interpretation, roll
motions seem to be virtually ubiquitous in the upper
convection zone and photosphere of the Sun. However, under solar
conditions, the hypothetical roll patterns are much more intricate in their
structure than the roll patterns observed in laboratory
experiments.

The outer diameters of the systems of concentric rings (``annular
rolls'') are of the order of the supergranular or mesogranular
sizes, so that such closed-ring systems could be an imprint of the
fine structure of the supergranular or mesogranular flows in the
subsurface layers.

The validity of our interpretation of the trenching patterns could be definitely stated only based on a comprehensive quantitative analysis. At the moment, the amount of information available is not yet sufficient for such an analysis and conclusion. Our aim here was merely to demonstrate, by characteristic examples, that the trenching patterns are widespread and diverse in their particular forms and that our algorithm can efficiently detect them. Nevertheless, we can now present here an illustration that appears to be an additional argument in favour of our interpretation.

We have averaged a series of images obtained in numerical simulations of granular-scaled solar convection by Rieutord \etal\ \cite{rieut}. The computations covered a domain that corresponded to a $30\times 30$~Mm$^2$ area of the solar photosphere. Each image contains $315\times 315$ pixels of size 95.24~km, and the interval between images corresponds to 20 seconds of real time. A two-hour-averaged image is shown in Figure~\ref{rieutproc}a, while Figures~\ref{rieutproc}b and \ref{rieutproc}c represent the results of processing the average using the COLIBRI code. It can easily be seen that, for widely-ranging $\Delta$ (from 8 to 16 pixels), the code does not highlight any clear-cut ridges or trenches. The resulting image is not sensitive to variations in $p$ over a range in which trenching was revealed in real solar images. It is thus fairly obvious that granular-sized polygonal convection cells cannot produce trenching patterns by themselves, and larger-scaled convection must be responsible for the arrangement of granules that results in trenching.

The local brightness of a time-averaged image of the solar
granulation reflects the local probability of the emergence of
granules. The light blotches in the image indicate the sites where
granules emerge most frequently. Qualitative reasoning (Getling,
2000) and correlation analyses (Getling, 2006) suggest that
granules may be overheated blobs entrained by the convective
circulation, and they can even repeatedly emerge on the solar
surface. In this case, granules appear as markers of roll convective flows, which
can thus be identified in averaged photospheric images.


\begin{acknowledgements}
      We are indebted to P.N. Brandt and R.A. Shine for making
      available the La Palma and SOHO MDI data, and to Th. Roudier for putting the results of numerical simulations at our disposal. We are also grateful to the referee and to L.M. Alekseeva
      for valuable comments on the manuscript. This work was
      supported by the Russian Foundation for Basic Research
      (projects 04-02-16580-a and 07-02-01094-a).
\end{acknowledgements}

\end{article}

\begin{thebibliography}{}

   \bibitem[1997]{baud}Baudin, F., Molowny-Horas, R., Koutchmy, S.: \aeta{1997}{326}{842}

   \bibitem[2004]{bdm-str}Berrilli, F., Del Moro, D.,
     Consolini, G., Pietropaolo, E., Duvall, T.L., Jr.,
     Kosovichev, A.G.: \sph{2004}{221}{33}

   \bibitem[2004]{bg}Brandt, P.N., Getling, A.V.: 2004,
     In Stepanov A.V., Benevolenskaya E.E., Kosovichev A.G.
     (Eds.), {\it Multi-Wavelength Investigations of Solar
     Activity, IAU Symp.} {\bf 223}, Cambridge Univ. Press, Cambridge, 231.

   \bibitem[1988]{dial}Dialetis, D., Macris, C., Muller, R., Prokakis, T.: \aeta{1988}{204}{275}

   \bibitem[1998]{book}Getling, A.V.: 1998, {\it Rayleigh--B\'enard
     Convection: Structures and Dynamics}, World Scientific, Singapore
     (Russian version: 1999, URSS, Moscow).

   \bibitem[2000]{getling}Getling, A.V.: 2000, {\it Astron. Zh.}
     {\bf 77}, 64 (English translation {\it Astron. Rep.} {\bf 44},
     56).

   \bibitem[2006]{g06}Getling, A.V.: \sph{2006}{239}{93}

   \bibitem[2002]{gb1}Getling, A.V., Brandt, P.N.:
       \aeta{2002}{382}{L5}

   \bibitem[2000]{hoekbr}Hoekzema, N.M., Brandt, P.N.: \aeta{2000}{353}{389}
   
   \bibitem[1998]{hoek3}Hoekzema, N.M., Brandt, P.N., Rutten, R.J.: \aeta{1998}{333}{322}

   \bibitem[2004]{lrt}Lisle, J.P., Rast, M.P., Toomre, J.:
       \aspj{2004}{608}{1167}

   \bibitem[1947]{mwh}Mann, H.B., Whitney, D.R.: 1947 {\it Ann. Math. Stat.} {\bf 18}, 50.

   \bibitem[1990]{mul}Muller, R., Roudier, Th., Vigneau, J.: \sph{1990}{126}{53}

   \bibitem[2002]{rast}
       Rast, M.P.: \aeta{2002}{392}{L13}

   \bibitem[2002]{rieut}Rieutord, M., Ludwig, H.-G., Roudier, T., Nordlund, \AA., Stein, R.: 2002,
     \textit{Nuovo Cimento} {\bf 25}, 523.

   \bibitem[2003]{roud5}Roudier, Th., Ligni\`eres, F., Rieutord, M., Brandt, P.N., Malherbe, J.M.: \aeta{2003}{409}{299}

   \bibitem[1997]{roud7}Roudier, T., Malherbe, J.M., November L., Vigneau, J., Coupinot, G., Lafon, M., Muller, R.: \aeta{1997}{320}{605}

   \bibitem[1997]{salov}Salov, G.I.: 1997, {\it Avtometriya},
     No. 3, 60 (English translation {\it Optoelectronics, Instrumentation and Data Processing}). 

   \bibitem[2000]{ssh}
     Shine, R.A., Simon, G.W., Hurlburt, N.E.:
     \sph{2000}{193}{313}

   \bibitem[1994]{simon}
     Simon, G.W., Brandt, P.N., November, L.J., Scharmer, G.B.,
     Shine, R.A.: 1994, In Rutten R.J., Schrijver C.J. (Eds.),
     {\it Solar Surface Magnetism, NATO Advanced Science
     Institute,} \textbf{433}, Kluwer Academic, Dordrecht,
     The Netherlands, 261.

   \bibitem[1951]{wh}Whitney, D.R.: 1951 {\it Ann. Math. Stat.} {\bf 22}, 274.

\end{thebibliography}
\end{document}